# Nonvolatile Static Random Access Memory (NV-SRAM) Using Magnetic Tunnel Junctions with Current-Induced Magnetization Switching Architecture

Shuu'ichirou Yamamoto[1,4] and Satoshi Sugahara[2,3,4]

[1]Department of Information Processing, Tokyo Institute of Technology, Yokohama 226-8502, Japan

[2]Imaging Science and Engineering Laboratory, Tokyo Institute of Technology, Yokohama 226-8503, Japan

[3]Department of Electronics and Applied Physics, Tokyo Institute of Technology, Yokohama 226-8502, Japan

[4]CREST, Japan Science and Technology Agency, Kawaguchi 332-0012, Japan

Abstract:

We propose and computationally analyze a nonvolatile static random access memory (NV-SRAM) cell using magnetic tunnel junctions (MTJs) with magnetic-field-free current-induced magnetization switching (CIMS) architecture. A pair of MTJs connected to the storage nodes of a standard SRAM cell with CIMS architecture enables fully electrical store and restore operations for nonvolatile logic information. The proposed NV-SRAM is expected to be a key component of next-generation power-gating logic systems with extremely low static-power dissipation.

1. Introduction

In recent years, power dissipation has been one of the most important concerns for highly integrated advanced complementary metal-oxide-semiconductor (CMOS) logic circuits, since it constrains their performance and the degree of device integration. In general, power dissipation in CMOS logic circuits can be divided into two components, dynamic and static power. The former is caused by on-current passing through CMOS logic gates due to logic operations, and the latter by leakage current in the CMOS gates during standby mode. The magnitude of the leakage current in CMOS gates is exponentially small in comparison with the on-current. However, static power dissipation gives rise to severe problems for CMOS logic circuits with very large scale integration. Recently proposed power-gating architecture based on multithreshold voltage CMOS (MTCMOS) technology is expected to be very effective at reducing static power dissipation in CMOS logic circuits.[1-7] In this type of architecture, logic circuits on a chip are partitioned into several circuitry domains, so-called power domains that are electrically separated from power supply lines and/or ground lines by sleep transistors, and these domains can be shut down during standby periods. Static power is thereby drastically reduced. A key technology for realizing power-gating systems is the backup of logic information in power domains. Static random access memory (SRAM) and its related latch circuits including a flip-flop (FF) play an essential role in data latches for high-speed logic circuits. However, these memory devices cannot be shut down without losing information. In addition, SRAM and FF dissipate static power during data retention owing to the presence of leakage current in their constituent CMOS gates. The recently developed embedded SRAM array in a microprocessor has the capability of entering "sleep" standby mode, thereby reducing the magnitude of supply voltage to the minimum value required for the memory cells to retain data.[3-6] However, this technique cannot reduce the static power to zero. One approach to realizing power-gating architecture for circuitry domains that require no power during standby is to transfer key information from the SRAM/FF in the power domains to a backup SRAM/FF placed in continuously powered circuitry before the power domains are shut down.[3,4,7] However, this architecture requires an inconvenient and complicated access procedure for data transfer through a signal bus line or a specially added one-to-one data transfer line. The limited data transfer rates of bus lines and the complex interconnected wiring of one-to-one data transfer lines are not suitable for power-gating systems with power domains that are frequently switched on and off. Another approach is to use nonvolatile memory such as ferroelectric random access memory (FeRAM) and magnetoresistive random access memory (MRAM) in power domains instead of SRAM and FF. However, these memory devices cannot easily replace SRAM and FF, since the operation speed of these devices cannot easily satisfy the requirements of SRAM and FF applications. Therefore, nonvolatile SRAM (NV-SRAM) and nonvolatile FF (NV-FF) with high performance comparable to that of standard SRAM and FF are indispensable for power-gating logic systems.

One of the most important applications of NV-SRAM and NV-FF is in nonvolatile power-gating processors as shown in Fig.1, which can be realized by applying NV-FF and NV-SRAM to important memory devices, such as the module configuration register, general purpose register, cache memory, and local storage memory. NV-SRAM and NV-FF can be shut down without losing data, and thus can enable efficient power-gating with domains that require no power during standby, leading to a drastic reduction of static power dissipation. Each power domain of the processor has the ability to execute a fast state transition between normal operation mode and "shutdown" mode, since NV-SRAM and NV-FF require no data transfer to backup memory devices. Using a power management unit to control the sleep transistors of the power domains, NV-SRAM and NV-FF make it possible to establish effective run-time power gating. Combined with a high-density nonvolatile memory (such as MRAM) as the main memory, a dynamic power-gating computing system can be established, as shown in the figure. It should be noted that the nonvolatile feature of the main memory is also beneficial, since it require no program/data loadings from external memory (such as a hard disk drive), when the computing system restarts to execute the suspended operation of a previous powered-on period.

NV-SRAM utilizing a standard SRAM cell with ferroelectric capacitors or resistive switching devices has already been demonstrated.[8-11] Magnetic tunnel junctions (MTJs) are an alternative promising candidate as a nonvolatile storage element for NV-SRAM. The Recent development of very high tunneling magnetoresistance (TMR) ratios at room temperature[12-14] is useful feature for MRAM[15-17] and would also be applicable to NV-SRAM. Although NV-SRAM cells using MTJs have already been reported,[18-20] they employ a magnetic field for magnetization switching of the MTJs, induced by current through the interconnections of the circuit. Thus, they potentially have serious problems for programming power, and programming disturbs (that is a technical term of failure mechanism for programming error). On the other hand, magnetic-field-free current-induced magnetization switching (CIMS) technology[21-26] allows fully electrical programming for MTJs, and it is expected that the problems of programming power and disturbs can be solved by CIMS technology. The circuit configuration of the above-described NV-SRAM cells using field-induced magnetization switching (FIMS)[18-20] cannot be applied to CIMS architecture, since MTJs in the cells using FIMS are placed in their inverter loop, and thus bidirectional current (which is required for CIMS from parallel to antiparallel magnetization and from antiparallel to parallel magnetization) is hardly generated.

In this paper, we propose and computationally analyze a new NV-SRAM cell using a pair of MTJs with CIMS architecture for fully electrical nonvolatile information storage.

2. Proposed NV-SRAM cell

Figure 2(a) shows the proposed NV-SRAM cell. The cell consists of two cross-coupled CMOS inverters with two pass transistors and a pair of MTJs. The cross-coupled inverters act as a bistable circuit; that is, a circuit that has two stable operating states that correspond to the logic information of 0 (a low voltage level, hereafter denoted by L) and 1 (a high voltage level, hereafter denoted by H). The pass transistors connected to the two storage nodes Q and /Q act as data-access transistors. The two MTJs are also connected to the two storage nodes. Logic information in the bistable circuit can be electrically stored as a magnetization configuration of these MTJs by CIMS using a pulse signal applied to the control (CTRL) line shown in Fig. 2(a), and thus the NV-SRAM cell can be powered off during shutdown mode without losing logic information. When the cell returns to the normal SRAM operation mode, the information stored in the MTJs is quickly restored to the corresponding logic state of the bistable circuit. This can be achieved as a result of the difference in the charging speed of the two storage nodes, owing to the asymmetry of resistances of the two MTJs. Note that NV-latch and NV-FF cells can be realized by connecting capacitive or resistive nonvolatile memory elements to the storage nodes of an inverter loop.[27-29] However, there have also not been yet any reports on NV-latch and NV-FF using MTJs with CIMS architecture.

NV-latch and NV-FF cells can also be easily configured in the same manner as the present NV-SRAM cell using a pair of MTJs. Figure 2(b) shows a positive edge triggered master-slave nonvolatile delay-FF (NV-DFF) that consists of a conventional latch and an NV-latch, as indicated in the figure.

3. Device modeling and simulation procedure

In order to analyze the circuit operation of the proposed NV-SRAM cell, we developed a SPICE (simulation program with integrated circuit emphasis) macromodel of an MTJ device including CIMS.[30] The fundamental circuit of our model consists of variable resistors and current-controlled ideal switches. The concept of the circuit model is similar to the model reported by Zhao et al.[31] However, our MTJ model can closely fit the electrical characteristics of recently developed CoFeB/MgO/CoFeB MTJs,[14] and includes CIMS phenomenon (discussed later). Figures 3(a) and 3(b) show the current-voltage and resistance-voltage characteristics of the developed MTJ model, respectively. The positive-current direction is defined by the direction from the free layer to the pinned layer in the MTJ model (denoted by f and p, respectively, in the inset of Fig. 3(a)). The junction resistances $R_{ap}(0)$ and $R_p(0)$ of the MTJ model in the antiparallel and parallel magnetization configurations at zero bias voltage were set to 16.7 and 8.33kΩ, respectively, resulting in a TMR ratio ($=[R_{ap}(0)-R_p(0)]/R_p(0)$) of 100%. This relatively high TMR ratio is easily achieved by recently developed MgO tunnel barrier technology.[12-14] MgO-based MTJs with a high TMR ratio showed nonlinear tunneling-type (Simons' formula type) current-voltage characteristics in the antiparallel magnetization configuration, and ohmic-like current-voltage characteristics in the parallel magnetization configuration.[14,21,22] Therefore, $R_{ap}$ and $R_p$ can be phenomenologically modeled as follows:

$$R_{ap}(V_{MTJ}) = R_p(0) + [R_{ap}(0) - R_p(0)] \cdot \exp\left(-\frac{|V_{MTJ}|}{V_{half}/\ln 2}\right), \quad (1)$$

$$R_p(V_{MTJ}) = R_p(0), \quad (2)$$

where $V_{MTJ}$ and $V_{half}$ are the bias voltage applied to the device and the $V_{MTJ}$ value when the TMR ratio is reduced to half its original value, respectively. This model can reproduce the finding that $R_{ap}$ is asymptotically close to $R_p$ at higher voltages. We compared our model with the experimentally obtained electrical characteristics of a MgO-based MTJ reported in ref. 14. Our model reproduced the experimental results within an absolute error of 1.5% and a standard error of 0.9%.

In the following simulations, a $V_{half}$ value of 0.5V was used[12] unless otherwise stated. The previously noted $R_p$ and $R_{ap}$ values were set so that currents through $MTJ_1$ and $MTJ_2$ exceed critical currents $I_{CF}$ and $I_{CR}$ at the desired bias voltage. $I_{CF}$ for CIMS from the antiparallel to parallel magnetization configuration and $I_{CR}$ from the parallel to antiparallel magnetization configuration were set to 30 and -30μA, respectively. (Their corresponding current density $J_C$ is $1\times10^6 A/cm^2$,[14,21] when the junction area is 0.003μm$^2$.) In general, the switching speed of CIMS depends on the current density through the MTJ. A lower current density causes a lower switching speed and requires a wider pulse duration. This phenomenon was included in our SPICE macromodel as a "switching delay". The switching delay is realized by timer circuitry, which consists of a voltage comparator and a delay circuit, and it can be set to the desired value by changing the resistance-capacitance product of the delay circuit. In this paper, the switching delay was set at a constant value of 1ns,[26] that is, shorter than the pulse width of the store signal applied to the MTJs through the CTRL line by a sufficient amount. In actual usages, the pulse width of the store signal may be prolonged to guarantee a successful store operation. In this paper, the sweep rate of the restore signal ($V_{SP}$) was set to 1V/ns. The rate should be chosen to assure successful restore operation, and a lower rate is preferable for achieving this. Note that a large store-signal pulse width and a low restore-signal sweep rate do not degrade the performance of normal SRAM operation.

Figure 3(c) shows the simulated response of the MTJ current when an alternating triangle voltage $V_{MTJ}$ is applied. CIMS hysteresis is clearly reproduced by our model without unwanted switching noise.

Operation of the proposed cell including its transition response was analyzed by HSPICE with a 0.07μm gate-length technology model.[32] The gate lengths(L)/widths(W) of n-channel and p-channel MOSFETs were set to 0.07/1μm and 0.07/1.5μm, respectively. The wide gates were used in order to readily achieve CIMS. Although the gate width can be reduced by optimizing the circuit design, a drastic reduction of the gate width requires a reduction of $J_C$ to achieve CIMS. This is the same situation as the case for MRAM using CIMS (spin RAM or spin transfer torque RAM).[21,22] When a relatively low $J_C$ value (such as $5\times10^5$A/cm$^2$)and a small MTJ (such as 0.001μm$^2$) are assumed, a narrow gate width with a W/L ratio of unity is applicable.[33]

The threshold voltages $V_{thp}$ and $V_{thn}$ of the p-channel and n-channel MOSFETs are -0.21 and 0.19V, respectively. The supply voltage $V_{SP}$ of the CMOS inverters in the cell during normal SRAM operation was set to 1V. In our simulation, the CTRL line is grounded, floating of the CTRL line or applying a small voltage to the CTRL line can also be appropriate to reduce unwanted leakage current through the MTJs of the NV-SRAM cell during normal SRAM operation. (A related discussion is presented in section 4.3.)

## 4. Results and discussion
### 4.1. Store operation of NV-SRAM

Operation of the NV-SRAM cell during normal SRAM operation proceeds in the same manner as an ordinary SRAM. The store operation can be performed only by applying a pulse signal to the CTRL line, when the bistable circuit is in one of the stable logic states. Figures 4(a) and 4(b) show schematic illustrations of the store operation of the proposed NV-SRAM cell, where the CTRL line is at levels L and H, respectively, and Fig. 4(c) shows taht of the restore operation (discussed later). In Figs. 4(a) and 4(b), the stored information of node Q is at level H and that of node /Q is at level L. The magnetization configuration of both $MTJ_1$ and $MTJ_2$ is antiparallel, which is realized by the operation of data reversal of the bistable circuit during normal SRAM operation, as shown in the last part of the time chart of Fig. 5. (Note that $R_{J2}$ ($R_{J1}$) shown in Fig. 5(a) (Fig. 5(b)) is less than $R_{ap}(0)$.) This is due to the bias dependence of $R_{ap}$ expressed by eq. (1). The store operation is independent of the initial magnetization configurations of $MTJ_1$ and $MTJ_2$, as described later. $I_{J1}$ and $I_{J2}$ in Fig. 4 represent currents through $MTJ_1$ and $MTJ_2$, respectively. The positive-current direction of $I_{J1}$ and $I_{J2}$ is the direction from the free layer to the pinned layer of $MTJ_1$ and $MTJ_2$, respectively, as described previously. It should be noted that in the parallel (antiparallel) magnetization configuration, a negative (positive) current induces CIMS but a positive (negative) current does not, as shown in Figs. 3(a) and 3(b). Figure 5(a) shows simulated waveforms of the NV-SRAM cell, where the time evolution of voltages on the WL, VSP, and CTRL lines, and of voltages $V_Q$ and $V_{/Q}$ at nodes Q and /Q, respectively, are plotted. Changes in the magnetization configurations of $MTJ_1$ and $MTJ_2$ are also respectively shown by $R_{J1}$ and $R_{J2}$ in the figure. Before the application of a signal to the CTRL line, negative $I_{J1}$ flows since the CTRL line is grounded and $V_Q$=H [Fig. 4(a)], and thus $R_{J1}$ remains at $R_{ap}$. $R_{J2}$ also maintains its initial value, since $I_{J2}$ cannot flow through $MTJ_2$ because $V_{/Q}$=L [Fig. 4(a)]. When a store signal on the CTRL line pulls up to level H, positive $I_{J2}$ can flow [see Fig. 4(b)] and $R_{J2}$ simultaneously decreases to $R_{ap}(V_{MTJ}=H)$, as shown in Fig. 5(a). Since $I_{J2}$ is designed to exceed $I_{CF}$ under this bias condition, the magnetization configuration of $MTJ_2$ switches to parallel after the switching delay, resulting in $R_{J2}=R_p$. In contrast, the magnetization configuration of $MTJ_1$ does not vary because $V_Q$=H, although the resistance $R_{J1}$ increases from $R_{ap}(V_{MTJ}=H)$ to $R_{ap}(V_{MTJ}=0)$ during the application of the CTRL signal, as shown in Fig. 5(a). As a result of the application of the store pulse to the CTRL line, the node information of $V_Q$=H is transferred into $MTJ_1$ as the antiparallel magnetization configuration ($R_{J1}=R_{ap}$), and the node information of $V_{/Q}$=L is transferred into $MTJ_2$ as the parallel magnetization configuration ($R_{J2}=R_p$). It should be noted that if $R_{J1}=R_p$ before the application of a store pulse, negative $I_{J1}$ that exceeds $I_{CR}$ causes CIMS, and thus $R_{J1}$ changes from $R_p$ to $R_{ap}$. However, if $R_{J2}=R_p$ in the initial state, positive $I_{J2}$ induced by a CTRL pulse does not cause CIMS, and thus $R_{J2}$ does not change. Therefore, the store operation is independent of the initial magnetization configurations of $MTJ_1$ and $MTJ_2$. When $V_Q$=L and $V_{/Q}$=H, these logic data are also transferred into $MTJ_1$ and $MTJ_2$ via the mechanism described above, as shown in Fig. 5(b).

4.2. Restore operation of NV-SRAM

The restore operation to return to a normal SRAM operation mode is also simple. The VSP line connected to $INV_1$ and $INV_2$ for power supply is pulled up, while the WL line is maintained at level L. Figure 4(c) schematically shows the charge and discharge current flows with respect to storage nodes Q and /Q. In this figure, $C_{s1}$, $C_{g1}$, $C_{s2}$, and $C_{g2}$ represent parasitic capacitances at nodes Q and /Q, which consist of wiring, gate-bulk overlap, and gate-source overlap capacitances in the p-channel and n-channel MOSFETs ($M_{L1}$, $M_{L2}$, $M_{D1}$, and $M_{D2}$) of $INV_1$ and $INV_2$. $V_Q$ and $V_{/Q}$ at storage nodes Q and /Q are given by

$$V_Q = \frac{C_{s1}}{C_{sg1}} V_{SP} + \int \frac{1}{C_{sg1}} (I_{L2} - I_{D2} - I_{J1}) dt, \quad (3)$$

$$V_{/Q} = \frac{C_{s2}}{C_{sg2}} V_{SP} + \int \frac{1}{C_{sg2}} (I_{L1} - I_{D1} - I_{J2}) dt, \quad (4)$$

where $C_{sg1} = C_{s1} + C_{g1}$ and $C_{sg2} = C_{s2} + C_{g2}$. $I_{L1}$ ($I_{L2}$) and $I_{D1}$ ($I_{D2}$) represent the charge and discharge currents with respect to node Q (/Q) through the p-channel and n-channel MOSFETs of $INV_1$ ($INV_2$), respectively. Simulated waveforms exhibited during the restore operation are also shown in Fig. 5(a), and the details of time evolution of $V_Q$, $V_{/Q}$, $V_{SP}-V_Q$, and $V_{SP}-V_{/Q}$ are shown in Fig. 6(a), where $V_{SP}$ increases from 0 to level H. Figure 6(b) shows the TMR ratio of $MTJ_1$ during the restore operation.

   In the initial stage of the restore operation (region I denoted in Fig. 6(a)), the charging of nodes Q and /Q is governed by the first terms of eqs. (3) and (4), and nodes Q and /Q are simultaneously discharged by $I_{J1}$ and $I_{J2}$, respectively. In this region, the charge and discharge currents $I_{L1}$, $I_{L2}$, $I_{D1}$, and $I_{D2}$ are negligible, because all the MOSFETs are in a cut-off condition. Since it can be assumed that $C_{s1} \approx C_{s2}$ and $C_{g1} \approx C_{g2}$ owing to the device and circuit symmetries and to the similar bias conditions of both inverters $INV_1$ and $INV_2$ during this stage, both $V_Q$ and $V_{/Q}$ gradually increase. However, $V_Q$ is always higher than $V_{/Q}$. This is due to the difference in the discharge rate between Q and /Q, which is because $R_{J1}(=R_{ap}) > R_{J2}(=R_p)$, and thus $I_{J1} < I_{J2}$. Note that although the TMR ratio of $MTJ_1$ decreases with increasing $V_Q$ because of its antiparallel magnetization configuration, the TMR ratio is not significantly degraded in region I. ($V_Q$ and the TMR ratio only increases to 0.04V and decreases to 95%, respectively, at the boundary between regions I and II shown in Fig. 6.) After $V_{SP}-V_{/Q}$ and $V_{SP}-V_Q$ exceed the threshold voltages $|V_{thp2}|$ and $|V_{thp1}|$ ($=|V_{thp2}|$) of $M_{L2}$ and $M_{L1}$, respectively, nodes Q and /Q are also charged by currents $I_{L1}$ and $I_{L2}$ derived from $M_{L1}$ and $M_{L2}$, respectively [region II in Fig. 6(a)]. During this stage, the operating points of $M_{L1}$ and $M_{L2}$ are in their saturation region, and thus $I_{L1}$ and $I_{L2}$ are governed by gate bias voltages $V_Q$ and $V_{/Q}$, respectively. Owing to the increase in $V_Q$, the TMR ratio of $MTJ_1$ is reduced to 77% at the boundary between regions II and III, as shown in Fig. 6(b). However, this TMR value is still sufficient to maintain the relation $I_{J1} < I_{J2}$. Therefore, $V_Q$ is higher than $V_{/Q}$. As a result, the relation $V_{SP}-V_{/Q} > V_{SP}-V_Q$ is maintained, which yields the relation of $I_{L2} > I_{L1}$. This accelerates the charging of node Q, and the voltage difference between $V_Q$ and $V_{/Q}$ is enhanced. Eventually, $V_Q$ becomes higher than the threshold voltage $V_{thn1}$ of $M_{D1}$. Thus, $M_{D1}$ is turned on, while $M_{L1}$ is turned off [region III in Fig. 6(a)]. Namely, node /Q is discharged by $M_{D1}$, and $V_{/Q}$ is further reduced. This leads to a further increase in the charging of $V_Q$ by $M_{L2}$, and thus $M_{D2}$ cannot be switched on. During this stage, the TMR ratio of $MTJ_1$ decreases to ~60%. However, the reduction of the TMR ratio has no serious impact on the restore operation, since the current drive capabilites of $INV_1$ and $INV_2$ become higher than those of $MTJ_1$ and $MTJ_2$ in region III, and an initial complementary push-pull operation of $INV_1$ and $INV_2$ is established; that is, $V_Q$ and $V_{/Q}$ are in a bistable condition governed by the two inverters, although $V_Q$ does not reach level H. Subsequently, $V_{/Q}$ rapidly approaches level L, and $V_Q$ increases to level H with increasing $V_{SP}$ [region IV in Fig. 6(a)]. The operating points of $M_{D1}$ and $M_{L2}$ enter their linear region. Finally, the feedback loop with the bistable condition of $V_Q \approx H$ ($V_Q$ is lower than level H owing to leakage current through $MTJ_1$ or $MTJ_2$, as discussed later) and $V_{/Q}=L$ is established, which correspond to the information stored in $MTJ_1$ and $MTJ_2$. After the restore operation, the normal mode of SRAM operation is possible.

### 4.3. Effect of MTJ on normal SRAM operation

A disadvantage of the proposed NV-SRAM cell is the flow of current between the level H storage node and the grounded CTRL line through $MTJ_1$ or $MTJ_2$ during normal SRAM mode operation. This leakage current results in excess power dissipation. The circuit current $I_{CELL}$ through the NV-SRAM cell can be defined by $I_{L1}+I_{L2}$ which include the leakage current induced by $MTJ_1$ or $MTJ_2$, is shown in Fig. 5(a). For the retention of data during the normal operation mode, the NV-SRAM cell requires an $I_{CELL}$ value of 87.6μA. Note that it decreases to 0.14μA when $MTJ_1$ and $MTJ_2$ are decoupled from nodes Q and /Q, respectively. The maximum value of $I_{CELL}$ is 451μA, which is caused by a normal SRAM operation of data reversal, owing to short circuit currents passing through the inverters. As shown in the last part of the time chart in Fig. 5, the operation of data reversal induces CIMS for one of the MTJs with the parallel magnetization configuration. As a result, both $R_{J1}$ and $R_{J2}$ develop the antiparallel magnetization configuration during normal operation mode (subsequent to the first reversal of data stored in the inverter loop). Obviously, a high $R_{ap}$ value can reduce the unwanted leakage current passing through $MTJ_1$ or $MTJ_2$, and a high $V_{half}$ value is also effective at reducing the leakage current. However, since an increase in $R_{ap}$ with a high $V_{half}$ value requires a high bias for CIMS, the $R_p$ and $V_{half}$ values must be carefully optimized in view of the trade-off between the leakage current in normal operation mode and the bias required for CIMS. Note that although the parallel magnetization configuration in $MTJ_1$ or $MTJ_2$ is established only during the period following the restore operation but prior to the first reversal of data, as shown in Fig. 5, this situation has no serious problems. Since the storage node connected to the MTJ with the parallel magnetization configuration is always at level L, no serious leakage current is generated.

The disadvantage of leakage current through $MTJ_1$ and $MTJ_2$ can be easily overcome by introducing selector transistors between the storage nodes (or CTRL line) and the MTJs, as shown in Fig. 7(a). The inverter loop of the NV-SRAM cell can be electrically separated from the MTJs by the selector transistors. Thus, the MTJs have no effect on the operation speed and the power dissipation during normal SRAM operation. Another interesting challenge is to utilize a pair of spin transistors, such as a pair of the spin-MOSFETs,[34-36] instead of the pair of MTJs, as shown in Fig. 7 (b). Their magnetization-configuration-dependent unique transistor behavior is also suitable for NV-SRAM and NV-FF operations. It is worth noting that the circuit configuration of a selector transistor with the MTJ shown in Fig. 7(a) acts as a spin transistor, referred to as a pseudo-spin-MOSFET,[37,38] in which its current drive capability can be modified by the magnetization configuration of the MTJ.

5. Conclusions

We propose and computationally analyze an NV-SRAM cell using MTJs with magnetic-field-free CIMS architecture. A pair of MTJs connected to the storage nodes of a standard SRAM cell with CIMS architecture make it possible to execute fully electrical store and restore operations for nonvolatile logic information. The proposed NV-SRAM technology using MTJs with CIMS can be applied to NV-latch and NV-FF, and these nonvolatile devices are expected to be important key components for power-gating logic systems of future high-performance and low-power integrated circuits.

Acknowledgment

The authors would like to thank Professors H. Maejima and H. Munekata of Tokyo Institute of Technology, Professor K. Inomata of National Institute for Materials Science, and Professor M. Tanaka of the University of Tokyo for fruitful discussions and encouragement. This work was partly supported by a Grant-in-Aid for Scientific Research from Ministry of Education, Culture, Sports, Science and Technology and the Core Research for evolutional Science and Technology (CREST) program of Japan Science and Technology Agency (JST).

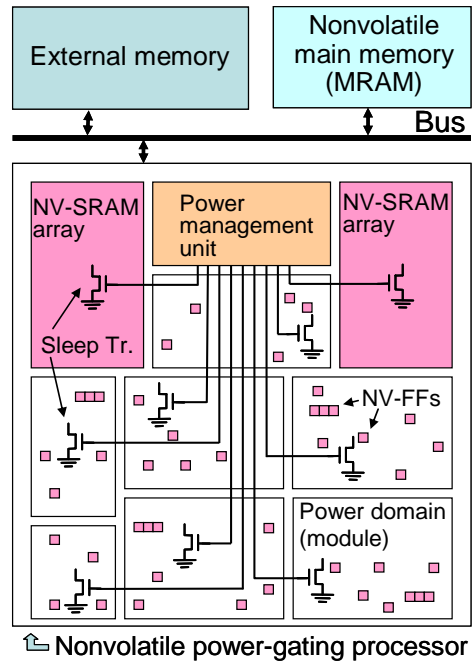

Fig. 1. Schematic illustration of nonvolatile power-gating processor. Combining nonvolatile main memory, a dynamically run-time power-gating computing system can be configured.

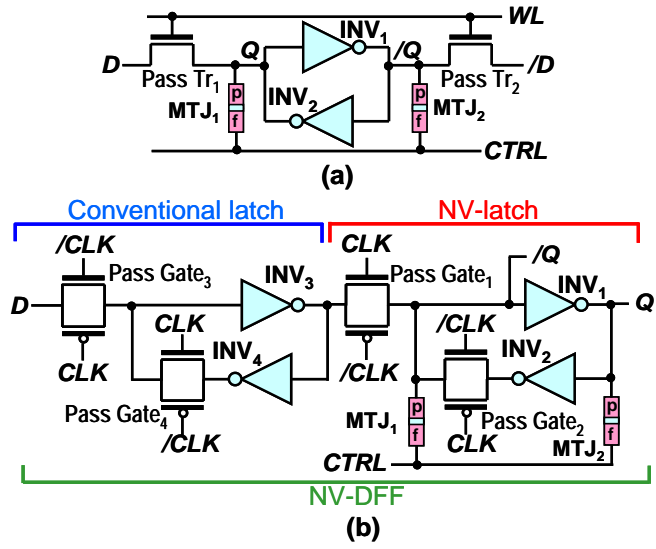

Fig. 2. Circuit configuration of the proposed (a) NV-SRAM cell and (b) positive edge triggered master-slave NV-DFF.

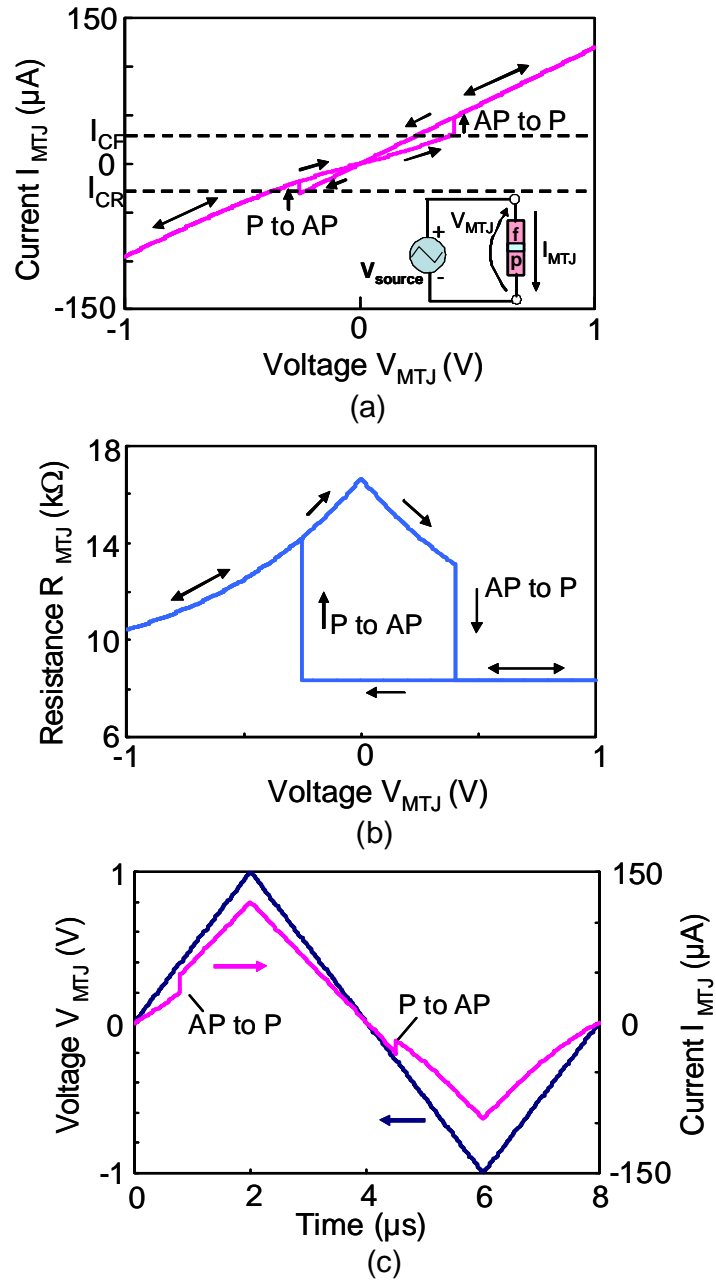

Fig. 3. (a) Current-voltage and (b) resistance-voltage characteristics of the present MTJ model. The current direction is shown in the inset, where f and p denote the free and pinned layers of the MTJ model, respectively. P and AP in the figure indicate the parallel and antiparallel magnetization configurations, respectively. (c) Simulated response of the MTJ current, when an alternating triangle voltage is applied to the MTJ model.

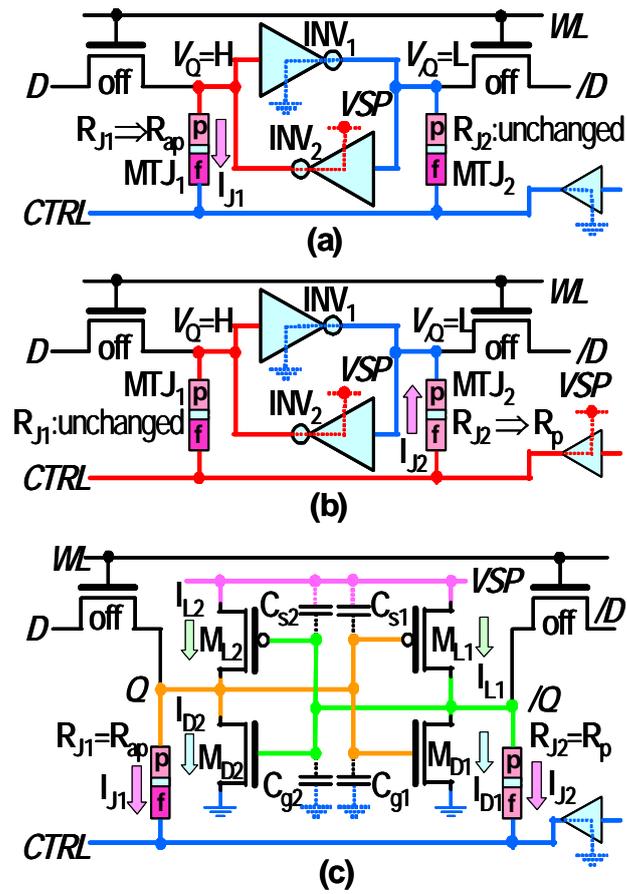

Fig. 4. Schematic illustrations of store operation with (a) level L and (b) level H voltages on the CTRL line. (c) Schematic illustration of restore operation, where $M_{L1}$ ($M_{L2}$) and $M_{D1}$ ($M_{D2}$) configure $INV_1$ ($INV_2$).

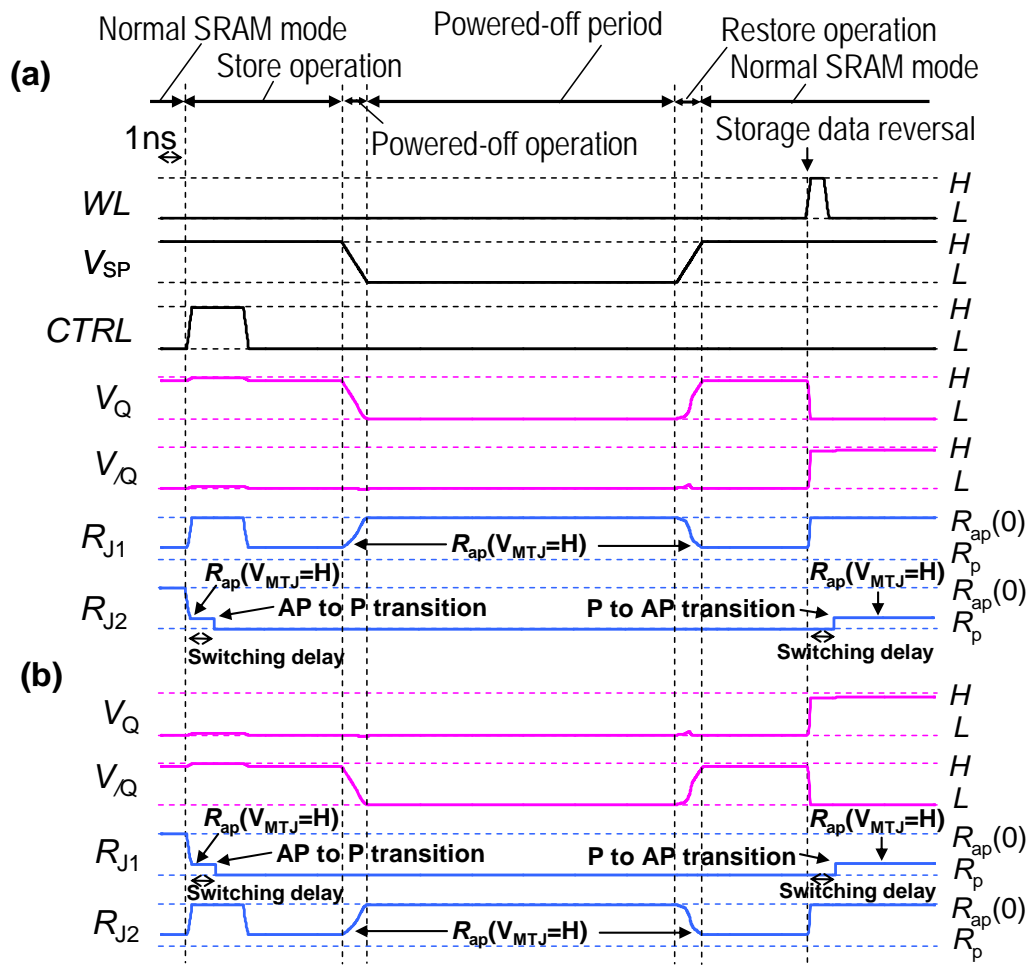

Fig. 5. Simulated input and output waveforms of the NV-SRAM cell for the store, restore, and normal SRAM mode operations with initial conditions of (a) $V_Q$=H and $V_{/Q}$=L and (b) $V_Q$=L and $V_{/Q}$=H.

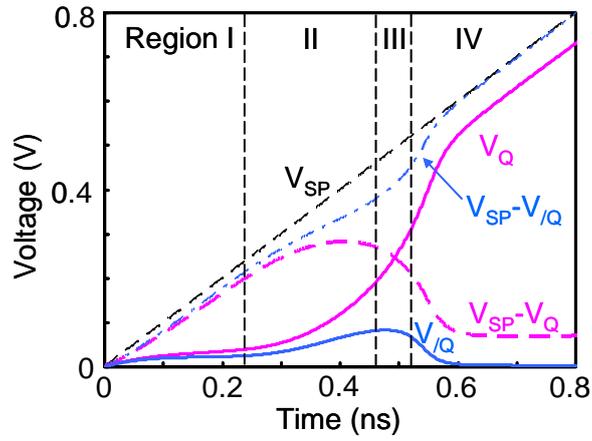

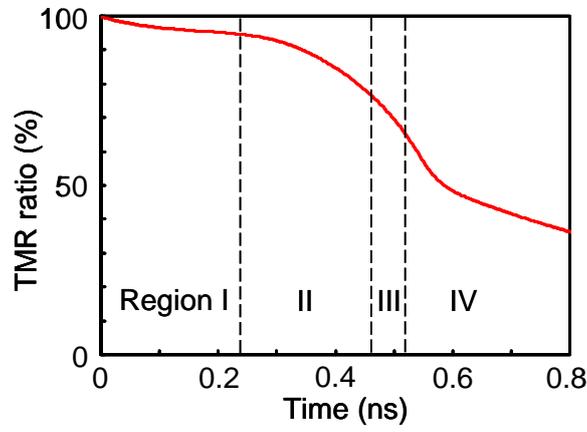

Fig. 6. (a) Time evolution of the node voltages $V_Q$ and $V_{/Q}$, and their differences with respect to $V_{SP}$. (b) TMR ratio of $MTJ_1$ during the restore operation.

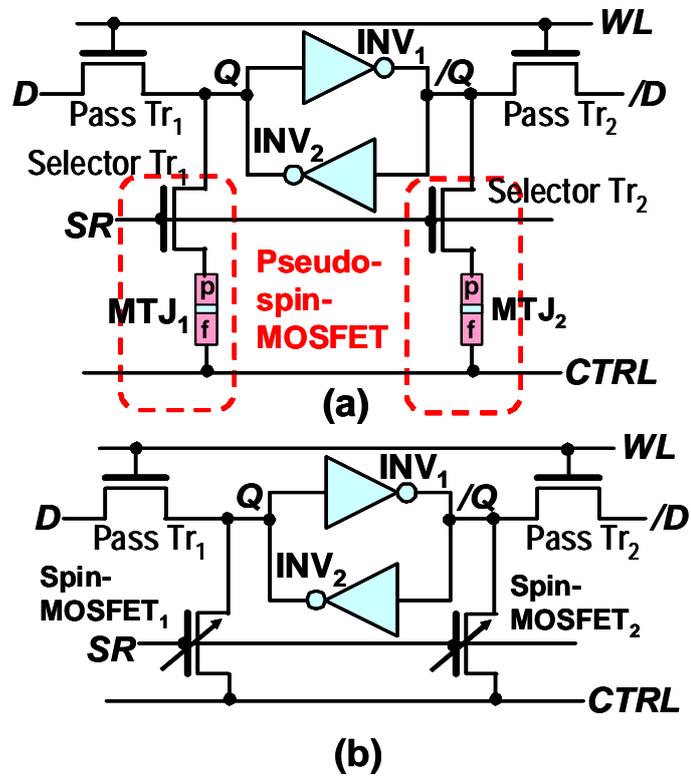

Fig. 7. Circuit configurations of NV-SRAM cells using (a) selector transistors and (b) spin-MOSFETs. The SR line is activated only in store and restore operation modes.